\documentclass[
    ,final            
  ]
  {aipproc}

\layoutstyle{6x9}

\usepackage{graphics}
\usepackage{psfig}
\usepackage{graphicx}
\usepackage{color}
\usepackage{dcolumn}

\newcommand{\ba}{\begin{eqnarray}}
\newcommand{\ea}{\end{eqnarray}}

\newcommand{\ov}{\overline}
\newcommand{\cd}{\! \cdot \!}
\newcommand{\be}{\begin{equation}}
\newcommand{\ee}{\end{equation}}
\newcommand{\bea}{\begin{eqnarray}}
\newcommand{\eea}{\end{eqnarray}}
\newcommand{\pa}{\partial}

\newcommand{\bw}{\begin{widetext}}
\newcommand{\ew}{\end{widetext}}

\begin{document}

\title{Brief introduction to viscosity in hadron physics
 \thanks{ Contribution to the Chiral10 Workshop. Valencia (Spain)}
}

\author{Antonio Dobado, Felipe J. Llanes-Estrada and Juan M. Torres-Rincon}{
  address={Departamento de F\'isica Te\'orica I, Universidad Complutense de Madrid, 28040 Madrid Spain}
}

\begin{abstract}
We introduce the concept of viscosity (both shear and bulk) in the context of hadron physics and in particular the meson gas, highlighting the current theoretical efforts to connect possible measurements of the viscosities to underlying physics such as a phase transition or the trace anomaly.
\end{abstract}

\maketitle

\section{Concept of viscosity}

The reader might be familiar with the Navier-Stokes equation, easily recognizable as a Newton's force equation for an element of viscous fluid
\be \label{NavierStokes}
\rho \left( \frac{\partial {\bf v}}{\partial t} +({\bf v}\cd {\bf \nabla}){\bf v}\right) = - {\bf \nabla} P +\eta \Delta {\bf v} +\left(\zeta + \frac{\eta}{3} \right) {\bf \nabla} ({\bf \nabla}\cd {\bf v}) \ .
\ee
The constant $\eta$ is the shear viscosity. It controls the flow of the momentum component $p_y$ perpendicular to the direction of flow $v_x$ as per
$j= - \eta \frac{\partial v_x}{\partial y}$. This is a first order transport equation, that assumes that the higher derivatives of the velocity are of negligible importance.
If the fluid is compressible, then a second transport coefficient, the bulk viscosity $\zeta$, controls the relaxation of longitudinal momentum gradients.

For a dilute gas, momentum transport is diffusive (a particle flows along carrying its momentum) and is hampered by the scattering cross section that interrupts the free streaming, for a non-relativistic hard-sphere gas $\eta = \frac{1}{3} n m \ov{v} \lambda$ in terms of the mean free path $\lambda=\frac{1}{n\sigma}$.
Thus, the larger the interaction, the smaller the viscosity; in such a diffusive system the ``perfect fluid''  and the ``perfect gas'' are opposite limits, characterized by strong the first and weak interactions the second, by low and high viscosity respectively (provided that the hydrodynamic description still applies).

Maybe it is interesting to note here that the dissipation of mechanical energy into heat is directly proportional to these viscosities
$$
\frac{dE_{mech}}{dt} = -\frac{\eta}{2} \int d{\bf x} \ (v_{i,k}+v_{k,i}-\frac{2}{3}\delta_{ik} {\bf \nabla}\cd{\bf v})^2 - \zeta\int d{\bf x} \ ({\bf \nabla}\cd{\bf v})^2
$$
(a temperature gradient adds a term proportional to the thermal conductivity $\kappa$, and so on for other transport coefficients).

To generalize the definition of viscosity to a relativistic system~\cite{Landaufluids} it is more convenient to look at the momentum-stress tensor than the force Eq.~(\ref{NavierStokes}). For a viscous fluid,
\be
\Pi_{ik} = \rho v_i v_k + P \delta_{ik}- \eta(v_{i,k}+v_{k,i}-\frac{2}{3}\delta_{ik}v^{l}_{\ .l}) - \zeta \delta_{ik}v^{l}_{\ .l} \ee
that can be carried over to special relativity with the substitution of the density $\rho$ by the enthalpy density $w$ because of the mass-energy equivalence, and the velocity $\bf v$ by the four-velocity $u_{\beta}$,
\be \label{Etensor}
T_{\alpha\beta} = -P \eta_{\alpha\beta} + wu_\alpha u_\beta + \tau_{\alpha \beta} 
\ee
with $\tau_{\alpha \beta}$ the dissipative part of the momentum-stress tensor.

This relativistic generalization suffers from a frame-definition ambiguity (choice of four-velocity $u_{\beta}$) that affects how dissipation is apportioned between thermal conductivity and bulk viscosity, for example. We follow Landau's convention that the proper frame is that in which the momentum flow vanishes, and the energy satisfies ideal hydrodynamics (no dissipation). This amounts to choosing $\tau_{\alpha\beta}u^\beta=0$ and lifts the ambiguity. Then the thermal conductivity does not appear in the dissipative part of the stress-energy tensor (it is confined to the particle-number dissipative flow)
\be \label{Etensor2}
\tau_{\alpha\beta}= -\eta(u_{\alpha ,\beta}+u_{\beta ,\alpha} + u_\beta
u^\gamma u_{\alpha ,\gamma}+u_\alpha u^\gamma u_{\beta ,\gamma} )
-\left(\zeta-\frac{2}{3}\eta\right)(\eta_{\alpha\beta}+u_\alpha u_\beta) u^\gamma_{\ \ ,\gamma} \ .
\ee

The fluid equations of motion are then the conservation law for the stress-energy tensor $T^{\alpha \gamma}_{\ \ \ \ ,\gamma}$ obtained by combining Eqs.~(\ref{Etensor}) and (\ref{Etensor2}).

\section{The KSS number}

In fluid mechanics it is common to construct dimensionless ratios that allow similarity analysis between different systems. Perhaps the best known of these is the Reynolds number $R\equiv \frac{\rho L u}{\eta} $ that quotients the density, characteristic fluid size and velocity, by the shear viscosity. High values of this ratio (low viscosities) imply turbulent, unstable flows, whereas small values (large viscosities) allow laminar, stationary flows.
These ratios however are contingent on the particular flow characteristics, and not on the thermodynamic state of a substance alone.
  
Recently, the ratio of viscosity to entropy density $\eta/s$ in natural units, or $k_B \eta/(\hbar s)$ in arbitrary units, has received much attention. This particular dimensionless ratio controls how fast shear perturbations damp, since the appropriate dissipative transverse dispersion relation is
$\omega(k) = -i \frac{\eta}{s} \frac{k^2}{T}$ (for vanishing chemical potential).

\begin{figure}
\psfig{figure=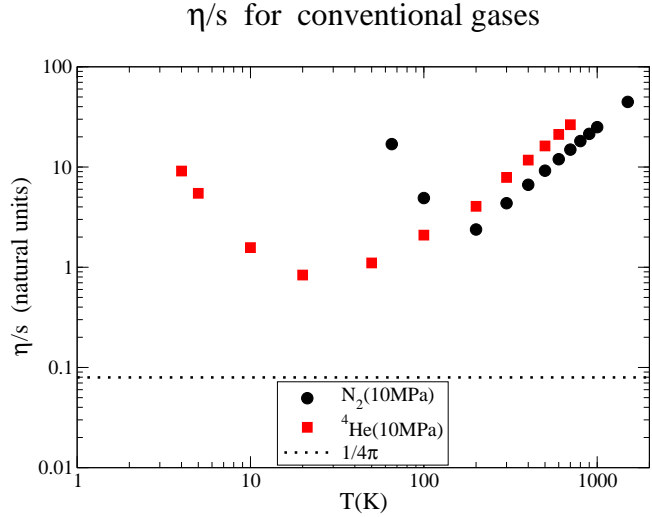,height=3.0in,angle=-90}
\caption{$\eta/s$ for molecular nitrogen and 	Helium gases together with
the KSS bound. \vspace{-0.3cm}
\label{normalgases}}
\end{figure}

The ratio $\eta/s$ has been much investigated because of the work of
Kovtun, Son and Starinets (KSS) \cite{Kovtun:2003wp}, who observed that in field theories accepting a gravity dual in higher dimension through the
holographic principle, the ratio could be estimated as a function
of the metric coefficients near a ``black brane''
$$
\frac{\eta}{s}= T f[g_{\alpha\beta}]
$$
and to their surprise, several feasible calculations with simple
metrics $g_{\alpha\beta}$ consistently yielded the value
$\eta/s=1/(4\pi)$. The field theories dual to these gravity
configurations are strongly coupled supersymmetric Yang-Mills
theories, far removed from our current physical picture of the
world. However, going to common substances whose viscosity and
entropy density values are tabulated, one finds (see Fig.
\ref{normalgases})
that this ratio is at least an order of magnitude larger than
$1/(4\pi)$. 		
The reader can find details in~\cite{Dobado:2007tm} about simple theoretical constructions that violate this bound, but the fact remains that no experimental system has been so observed.

An intuitive argument~\cite{Danielewicz:1984ww} for the existence of such a bound is given by Heisenberg's uncertainty principle, $\ov{\epsilon}\tau\geq 1$, with $\tau$ the characteristic damping
time (controled as above by the viscosity), and $\ov{\epsilon}$ the average energy per particle in the plasma. Some dimensional analysis transforms this relation in the KSS bound, excepting
 the $1/(4\pi)$ prefactor.

The ratio can also be approximately extracted in Relativistic Heavy Ion Collisions, and is there shown to be very close to the lower bound, which has lead to the expression ``perfect liquid'' to refer to the strongly interacting nuclear plasma (perhaps describable in terms of quarks and gluons instead of hadrons). 
The key observable is the so called ``elliptic flow''. To understand it, 
the reader just needs to return to Eq.~(\ref{NavierStokes}) and start by looking at the ideal-fluid part of the equation. A larger gradient of pressure translates into a larger acceleration, while the viscosity terms damp this additional acceleration.

In Heavy Ion Collisions the nuclei do not always collide head-on, but only part of the two spheres overlap. Then the pressure gradient in the reaction plane is larger than the pressure gradient in the perpendicular direction. 
The resulting momentum distribution of emitted hadrons peaks therefore at higher momenta for in-plane hadrons causing an azimuthal anisotropy around the beam axis. Its clear observation indicates that the viscosity is low.

\section{Microscopic computations of viscosities}
The one-particle out-of-equilibrium distribution function $f(\mathbf{x},\mathbf{v},t)$ satisfies a transport equation
\be \frac{df(\mathbf{x},\mathbf{v},t)}{dt} = C [f(\mathbf{x},\mathbf{v},t)]\ee
or more explicitly, for a boson gas,
\ba 
\label{BUU} \frac{\pa f}{\pa t} + \frac{\mathbf{p}}{E(p)} \mathbf{\nabla} f = \int d \overline{\sigma} d \mathbf{p}_1 v_{\textrm{rel}} \\ \nonumber
\left[f' f'_1 \left( 1+ \frac{(2 \pi)^3}{N} f\right) \left( 1+ \frac{(2\pi)^3}{N}f_1\right) - f f_1 \left( 1+ \frac{(2 \pi)^3}{N} f'\right) \left( 1+ \frac{(2 \pi)^3}{N} f'_1\right) \right].
\ea

Slightly out of equilibrium, $f \simeq  f_0 + \delta f$,
where $f_0$ is the Bose-Einstein distribution function. The perturbation function $\delta f$ is proportional to the gradients of the hydrodynamical fields out of equilibrium, the velocity if one deals with viscosities. To obtain the shear viscosity we parametrize $\delta f$ as~\cite{Dobado:2003wr}

\be \label{devoutofeq}
f = f_0 \left[1-\frac{g_\eta(p)}{T} \Delta_{ij} \tilde{V}^{ij} \right], 
\ee

where $\Delta_{ij} \equiv p_ip_j  -\frac{1}{3} \delta_{ij} p^2 $ and $\tilde{V}_{ij}$ represents the gradient of the velocity, $v_i$ :
\be \tilde{V}_{ij} = \frac{1}{2} (\pa_i v_j + \pa_j v_i) - \frac{1}{3} \pa_k v^k \delta_{ij}.
\ee

Eq.~(\ref{BUU}) is a slightly involved integral equation that is solved projecting it into an appropriate basis of functions, performing the multidimensional integral by Montecarlo, and then inverting the matrix truncation of the collision operator $C$.

Once solved, the shear viscosity, a macroscopic hydrodynamic coefficient, is expressed as an integral over the distribution function, dependent on such microscopic details as the mass of the particles in the gas and the cross section between them, and a function of the temperature.

An analogous derivation yields the bulk viscosity, except Eq.~(\ref{devoutofeq}) is substituted by
\be
f=f_0\left[1-\frac{g_\zeta(p)}{T} {\bf \nabla}\cd{\bf v}\right]
\ee
and we can compare them
\ba 
\eta & = &  \frac{1}{10T} \  \int  \frac{d\mathbf{p}}{E(p)} f_0 p_i p_j \Delta^{ij} g_\eta(p), \\ \nonumber
\zeta & = & \frac{1}{T} \  \int  \frac{d\mathbf{p}}{E(p)} f_0  
\left( \frac{p^2}{3}-v_s^2 E(p)^2\right) g_\zeta(p),
\ea
(having introduced the speed of sound $v_s$).

The computation of the bulk viscosity is complicated by the presence of zero modes in the corresponding collision operator $C$. If one substitutes $g_{\zeta}=1$, the symmetry properties of $C$ yield zero. This zero mode is associated to particle number conservation and is not really a departure from equilibrium. Associated to energy conservation there is a second zero mode $E$. 

Therefore one has to exclude the constant and linear $\propto E$ functions from $\delta f$ for bulk distortions, and invert the collision operator in the perpendicular space to these zero modes.

The complications with the bulk viscosity are further enhanced by the Landau-Lifschitz condition, that becomes a restriction on the allowed $g_\zeta$, given by
$\int  \frac{d\mathbf{p}}{E(p)} f_0 E(p)^2 g_\zeta(p) =0$.

\section{Shear viscosity and phase transitions}
It was stressed in~\cite{Csernai:2006zz} that the shear viscosity has a minimum near phase transitions in conventional fluids as in Fig.~\ref{normalgases}, and that this could be used in
experimental work to pinpoint the precise location of the possible phase transition between the hadron gas and the quark-gluon ``perfect liquid''. Detailed work on the pion gas is consistent
with this hypothesis~\cite{Chen:2007xe,Dobado:2008vt}.

Since there is no formal proof of this statement, but just indications from particular calculations in QCD, we have studied the issue in a toy model that allows for a much more reliable treatment
(still, in mean-field approximation for the vacuum condensate), the linear $\sigma$ model in large-$N$~\cite{Dobado:2009ek}. This model is the counterexample that, although the phase transition and
 the minimum of viscosity over entropy are related, the temperature of both do not need to coincide.

\begin{figure}
\includegraphics[height=2.5in]{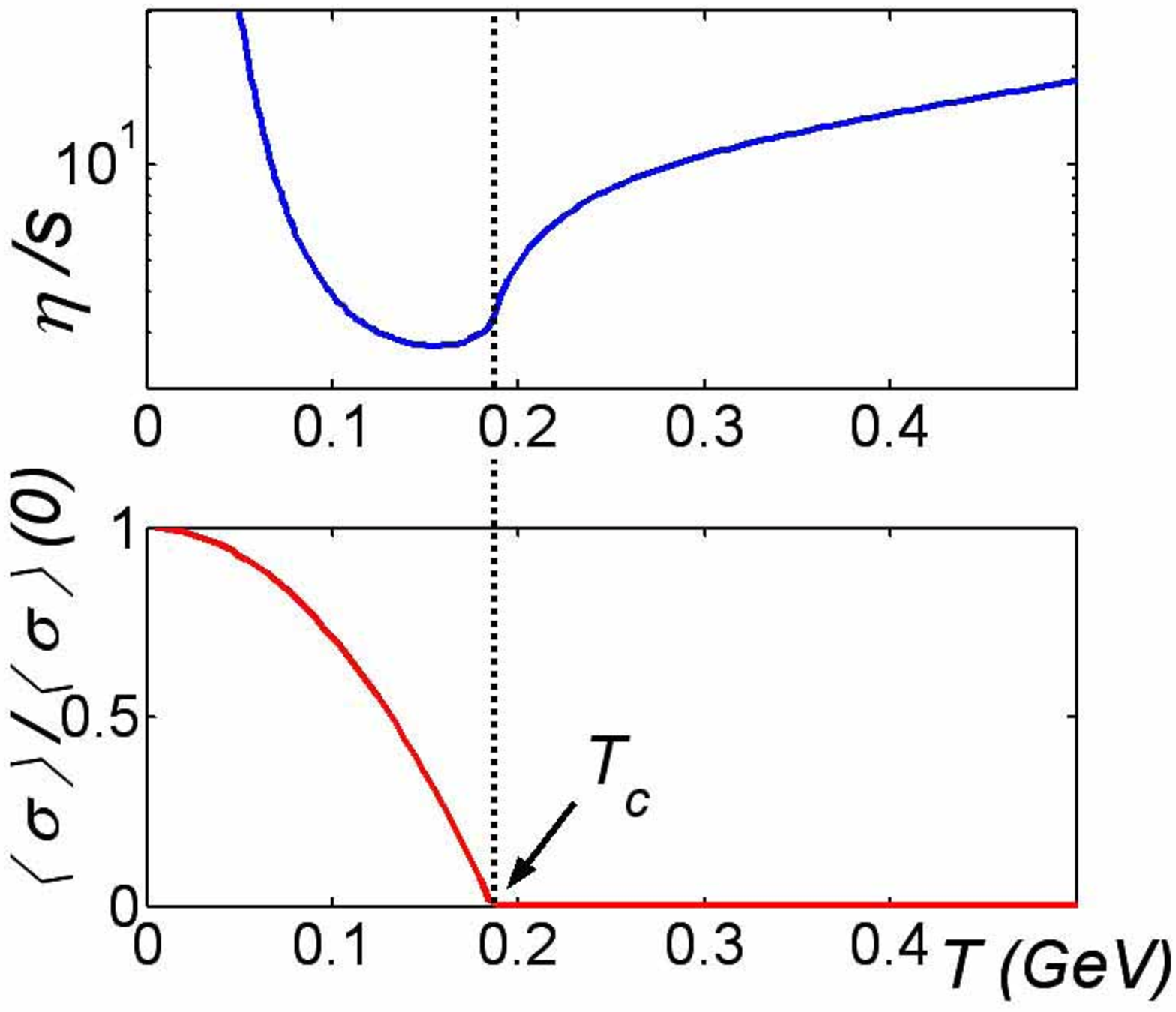}
\caption{The minimum of $\eta/s$ occurs just before the critical temperature for the phase transition in the Linear Sigma Model. This is where the condensate varies rapidly approaching zero. The second order phase
 transition is marked by a non--analyticity of the ratio.
\label{fig:minimumTc}}  
\end{figure}

The situation is clearly seen in Fig.~\ref{fig:minimumTc}.
The KSS number is seen to have a non-analiticity at the second-order phase transition where the condensate vanishes, but the minimum is seen to be reached distinctly before this phase transition, when
 the absolute value is dropping and affecting the cross section for pion scattering (and hence the shear viscosity) most greatly.

In conclusion, although true that viscosity over entropy density is an indication of the phase transition, we do not dare conclude that it will pinpoint the phase transition temperature precisely.

\section{Bulk viscosity and scale invariance}

The bulk viscosity is sensitive to the conformal breaking of the system. If the system is dilatationally invariant, then $\zeta$ is equal to zero. If conformal
symmetry is broken (e.g. by the explicit presence of dimensionful parameters in the Lagrangian like a mass term) then $\zeta$ turns out to be nonzero and proportional
to this breaking term. The conformal invariance can also be broken by pure quantum effects (trace anomaly), where
the interaction measure (defined as the trace of the momentum-stress tensor normalized by the temperature $T^{\alpha}_{\ \ \alpha}/T^4$) is nonzero.
This growth of the trace anomaly near $T_c$ has been proposed to drive a maximum in the bulk viscosity.

\subsection{Bulk viscosity of the pion gas}

\begin{figure}
  \includegraphics[height=.3\textheight]{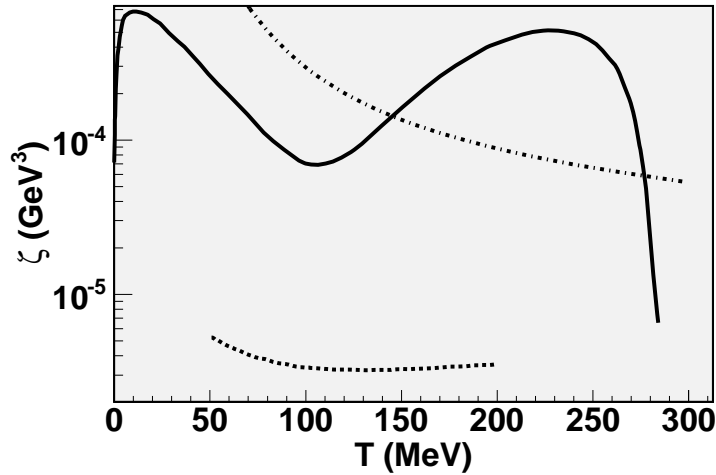}
  \caption{The value of the bulk viscosity in the pion gas, as well as its temperature dependence, has not yet been settled. Solid line: chiral perturbation theory based calculation \cite{FernandezFraile:2008vu}.
Dashed line: kinetic theory calculation based on phase shifts \cite{Davesne:1995ms}. Dashed-dotted line: Our preliminary result in the independent particle approximation for the pion gas. \label{figDani}}
\end{figure}

In the pion gas the relation between the trace anomaly and the bulk viscosity has been presented in \cite{FernandezFraile:2008vu} showing a peak in $\zeta$ when the effects
of the trace anomaly at $\mathcal{O} (T^8)$ in the chiral Lagrangian are included. However, in the recent work \cite{FernandezFraile:2010gu} where the bulk viscosity in the massive Gross-Neveu model
in $1+1$ dimensions is calculated, it is shown that there is no clear relation between the maximum in the interaction measure and a peak in the bulk viscosity, showing the latter
a monotonous behaviour with the temperature. Moreover, some of the previous results on the bulk viscosity are not always compatible as we can see in the Fig. \ref{figDani} and show
important quantitative discrepances.

To clarify this situation we are calculating the bulk viscosity in the pion gas with unitarized chiral perturbation theory in the interactions both the elastic collision process and the speed of sound.
In the low temperature limit (where the inelastic processes that enter are exponentially suppressed), the bulk viscosity asymptotically behaves as
\be \zeta \sim \frac{f^4_{\pi} \ T^{-7/2}}{m_{\pi}^{11/2}} \ee
and to zero at high temperature. In the independent particle approximation \cite{Arnold:2006fz}, where the thermodynamics is that of an ideal gas,
this occurs monotonously. The speed of sound has a universal behaviour in this approximation, monotonously increasing towards the aymptotic value of $1/\sqrt{3}$.

Our preliminary result within the independent particle approximation to Eq. \ref{BUU} is shown in Fig. \ref{figDani}. The order of magnitude of $\zeta$ is compatible with the previous result
 of \cite{FernandezFraile:2008vu}, but the qualitative behaviour is strictly monotonous in agreement with \cite{Davesne:1995ms}, presumably because of the universality of the speed of sound.
 The peak around $T=230$ MeV in \cite{FernandezFraile:2008vu} could appear in our calculation by improving the dependence of $c_s$ on temperature.

We are currently investigating these discrepancies in detail, as well as lifting the independent particle approximation including the effect of a more realistic behaviour of the speed of sound
through the unitarized cross section coming from the Inverse Amplitude Method applied to the chiral perturbation theory. The results will be published elsewhere.

\begin{theacknowledgments}
Work supported by grants FPA 2008-00592, FIS2008-01323 plus 227431, HadronPhysics2 (EU) and PR34-1856-BSCH, UCM-BSCH GR58/08, 910309, PR34/07-15875. JMT is a recipient of an FPU scholarship.
\end{theacknowledgments}


\bibliographystyle{aipproc}   

\end{document}